\documentstyle[12pt]{article}
\newfont{\larom}{cmbx10 scaled\magstep3}
\newfont{\bsan}{cmssbx10}
\newfont{\bfit}{cmbxti10 scaled\magstep1}
\begin{document}
\begin{center}
  {\larom Dogmatism and Theoretical Pluralism in Modern Cosmology}

  \vspace{10mm}
  {\large \it Marcelo B.\ Ribeiro$^{\dag}$\footnote{ \ {\bf Present
  address:} Department of Mathematical Physics, Institute of Physics, Federal
  University of Rio de Janeiro, C.P.\ 68528, Ilha do Fund\~{a}o, Rio de
  Janeiro, RJ 21945-970, Brazil; e-mail: mbr@if.ufrj.br} 
  and Antonio A.\ P.\ Videira$^{\dag \: \S }$}

  \vspace{10mm}

\end{center}
\begin{flushleft}
  {\normalsize \hspace{1.95cm} $^{\dag}$ Observat\'{o}rio Nacional--CNPq,
  Rio de Janeiro, Brazil}

  {\normalsize \hspace{1.95cm} $^{\S}$ Department of Philosophy,
  State University of Rio de Janeiro, Rua \\ \hspace{2.25cm}
  S\~{a}o Francisco Xavier 524, Rio de Janeiro, RJ 20550-013, Brazil;
  \\ \hspace{2.25cm} e-mail: guto@on.br}

\end{flushleft}
\begin{center}

  \vspace{9mm}

\end{center}

\vspace{9mm}

SUMMARY

\begin{quotation}
  \small
  This work discusses the presence of a dogmatic tendency within
  modern cosmology, and some ideas capable of neutralizing its negative
  influence. It is verified that warnings about the dangers of dogmatic
  thinking in cosmology can be found as early as the 1930's, and we
  discuss the modern appearance of ``scientific dogmatism''. The
  solution proposed to counteract such an influence, which 
  is capable of neutralizing this dogmatic tendency, has its origins
  in the philosophical thinking of the Austrian physicist Ludwig
  Boltzmann (1844-1906). In particular we use his two main
  epistemological theses, scientific theories as representations of
  nature and theoretical pluralism, to show that once they are embodied
  in the research practice of modern cosmology, there is no longer
  any reason for dogmatic behaviours.
\end{quotation}
\vspace{6mm}
\newpage
\vspace{5mm}
\begin{flushright}
{\it ``In cosmological studies, then, a knowledge of the history and \\
philosophy of science is not a superfluity, it is a necessity.''}

\vspace{5mm}

H.\ Dingle \cite{d}
\end{flushright}

\section{\protect\normalsize \bf INTRODUCTION}

In a recently published article, Matravers, Ellis and Stoeger 
\cite{mes} stressed that  the development of cosmology as a
scientific discipline requires that cosmological models other
than the standard Friedmann-Lema\^{\i}tre-Robertson-Walker (FLRW)
models be considered as effective alternatives to the standard
cosmological scenario. They stated the thesis that the intrinsic
power of those alternative models lies in the fact that, by
approaching the cosmological problem from a more descriptive and
observationally oriented perspective, as opposed to the standard
view which starts from broad explanatory premises, one could
observationally justify the latter. In this way those
phenomenologically oriented cosmologies allow the possible empirical
confirmation, or denial, of the basis of the FLRW models.

Matravers et al.\ are aware of the fact that, so far, those
complementary approaches, including the one they specifically 
propose in the article, still lack the comprehensive explanatory
power achieved by the standard model, and, therefore, considerable
work still needs to be done in order to fill the gap between these
two views, or even to implement some of the steps outlined by
the alternative approach to cosmology. Hence, they do not regard
the standard model as outdated. On the contrary, they wish to keep
it in a healthy contact with observational issues and to develop
a more self-critical and less doctrinaire cosmology where the two
complementary approaches to cosmology can interact in a mutually
beneficial way \cite[p.\ 31]{mes}.

However, it is from within the community favorable only to the standard
FLRW model that Matravers et al.\ indicate the existence of a strong
resistance in recognizing and allowing that the theoretical work is
vulnerable to observational falsification, pointing out that such a
resistance is especially prevalent among the researchers favorable
to the inflationary scenario \cite[p.\ 35]{mes}. Their paper can,
therefore, be seen as an attempt to counteract what can only be
described as a latent dogmatic tendency coming from within the
cosmological community. For this reason, they also stress the need
for a more self-critical attitude from researchers of the field
in general. In their words: 
\begin{quotation}
 \small \sf
 \noindent In fact, a rather serious and disturbing situation has 
 developed within modern cosmology, in which some workers promote certain
 cosmological theories as correct and well-established without seeming to
 regard the adequacy of their observational or experimental justification
 as of any importance. At the same time they tend to dismiss more
 observationally-based approaches -- for example the kind of larger
 justificatory investigation we have just proposed -- as being unnecessary
 or even `unscientific', simply because such approaches do not
 unquestioningly incorporate the standard view. This attitude is itself
 dangerously close to being unscientific, for it elevates theory above
 observation and relies on simplified geometrical models (certainly of
 considerable explanatory power) without subjecting them to adequate
 observational testing -- or even denying that they should be tested
 \cite[p.\ 31]{mes}.
\end{quotation}

We agree with Matravers et al.\ that a rather dogmatic and
dangerously unscientific attitude has developed in modern
cosmology \cite[p.\ 35]{mes}. Other authors like Tolman \cite{tolman},
MacCallum \cite{mm1}, Wesson \cite{wesson},
Rothman and Ellis \cite{re}, Krasi\'nski \cite{k} and, especially,
de Vaucouleurs \cite{vau}, have also expressed similar views. They
all warned of the danger of strongly believing in ideas not confirmed by
observations, pointing out that without this confirmation we lose the only
way we can distinguish science from metaphysics. For instance,
de Vaucouleurs was very clear about this point: 

\begin{quotation}
 \small \sf
 \noindent Unfortunately, a study of the history of modern cosmology
 (\ldots) reveals disturbing parallelisms between modern cosmology
 and medieval scholasticism. (\ldots) Above all I am concerned by
 an apparent loss of contact with empirical evidence and observational
 facts, and, worse, by a deliberate refusal on the part of some
 theorists to accept such results when they appear to be in conflict
 with some of the present oversimplified and therefore intellectually
 appealing theories of the  universe. (\ldots) [That concern] is due
 to a more basic distrust of doctrines that frequently seem to be more
 concerned with the fictitious properties of ideal (and therefore
 nonexistent) universes than with the actual world revealed by
 observations \cite{vau}.
\end{quotation}

The few quotations and references above do not indicate a widespread
presence of dogmatism among cosmologists, but simply its presence
among at least some of them. However, the fact that at different
times, different people possessing different theoretical and
observational perspectives and motivations, not only acknowledged
the presence of dogmatism in cosmological research, but were also
worried about its influence, is enough to show that such an influence
is not negligible and should be investigated. Therefore, it is
reasonable to say that dogmatic tendencies have been felt in the
field since at least the 1930's.

Nevertheless, it is clear that those dogmatic tendencies within
modern cosmology have little to do with the FLRW model itself,
whose achievements so far can only be described as impressive.
They result from the {\it attitude} of not accepting that some
key features derived from those models can be checked, let alone
questioned, by observations. When a feature of a model is ascertained
through imposition rather than by experimental or observational
check it is unscientific because it is {\it only} based on personal
choices. In other words, a certainty achieved that way becomes
a dogma.

Here we shall discuss the issue of dogmatism in modern cosmology.
We accept the quotations and references above as sufficiently
enough evidence of its presence in modern cosmology,
and we will propose a way capable of neutralizing its influence.
In our opinion, that can be achieved if researchers in the field
embody the epistemological principles advanced by Ludwig Boltzmann
(1844-1906) from the end of the 19th century until his death.

\section{\protect\normalsize \bf DOGMATISM IN MODERN PHYSICS}

It is generally accepted that scientific truth is achieved
when theory is directly confronted with observations (or
experiments). Since the time of Galileo Galilei (1564-1642),
observation and/or experimentation has been used to confirm
or to falsify a theory, and without such crucial tests no
theory can be considered scientific. To accept a theory
without this experimental/observational validation is to
accept it as a dogma.

However, things are not so simple when such a validation is not
clear cut and free from ambiguities, which in practice is the
case most of the time when doing real science. In such
situations dogmatic tendencies can thrive. Those tendencies
appear in the history of physics more frequently than one 
may imagine. An example particularly important to the subject
discussed in this article was the debate between Boltzmann and
Wilhelm Ostwald (1853-1932) at the end of last century concerning
the atomic view of the world. At stake was the definition of a
scientific theory, what should be its aims and methods, and 
the definition of scientific truth.

\subsection{\protect\bfit Boltzmann and the dogmatism in
	    Physics at the end of the last century}

By the end of the last century, Boltzmann was engaged in a
passionate defense of the atomic concept which, at the time,
was facing a growing number of powerful opponents, like
Ostwald and Georg Helm (1851-1923), who considered the atomic picture
of the world outdated \cite{B,BB,B1,BB1,o}, \cite[p.\ 42-61]{guto}. They
then advocated its replacement by the view that the physical world
could only be correctly described by means of the concept of
energy conservation and its derivatives, which implied the denial
of the atomic idea \cite{o,debate}.

Boltzmann feared that such a purely energetic representation
would lead physics to become dogmatic, a fact that would inevitably
also lead to its stagnation. He then wrote many epistemological texts
about the development of physics in general, whose conclusions
led him to advance what is now considered his two main
epistemological theses \cite{guto,debate}. The first one stated
that a physical theory is nothing more than a representation of
Nature, and the second thesis stated that Nature can be represented
by many different theories, which can even be opposed to each other.
Nowadays this last thesis is known as {\it theoretical pluralism}.

Among the physicists of the last one hundred and fifty years,
Boltzmann was one of the few, if not the only one from within
the scientific community and, therefore, from the perspective
of an active and eminent physicist, to discuss dogmatism
in an epistemological context. His epistemological thinking covers
issues like what a scientific theory is and how it develops. For
those reasons, we believe that his ideas give us the appropriate
epistemological framework which allows us to identify and
counteract  what can inhibits the development of scientific
theories, namely dogmatism.

Nevertheless, one may ask the question: how is it possible that
dogmatic tendencies can thrive even when the scientific community
openly accepts that the ultimate test of a theory is 
experimentation? We shall discuss this point next.

\subsection{\protect\bfit What is ``scientific'' dogmatism?}

It is generally accepted nowadays that in science nothing is in
principle unquestionable, but, inasmuch as the validation of new
theories and models usually takes time, a certain degree of
conservatism towards new theories and models, and skepticism
towards new observations, is, nevertheless, necessary since it
is not possible to build a sound conceptual and experimental scientific
body when there is a continual change in the fundamental scientific
concepts. Such skepticism is also evidence of the existence of
critique in science, which is one of the most important ingredients of modern
scientific reasoning and practice. Therefore, orthodoxy plays the
healthy role of preserving the scientific knowledge obtained on
solid bases until new theories prove to have sufficient internal
consistency and experimental validation.

However, when strong conservatism and orthodoxy becomes deep
rooted in the scientific community, a situation may arise
that, if not effectively and successfully challenged, may lead
the community to avoid altogether any kind of change of the
established ideas. Such a rejection to change may easily turn
to aggression towards the proposers of new ideas. In such an
environment the established theories crystallize, becoming
dogmatic, and scientific debate ceases to exist.

Such strong conservatism and orthodoxy very frequently come
about when researchers mistakenly take their theories to be
the researched objects themselves, believing that the former
coincide with the latter. By doing this, they identify theory
with object, and in this identification it is implicitly or
explicitly assumed that the role of a scientific theory is to
bring to our knowledge Nature itself. Therefore, for those
researchers scientific truth means exactly such identification.

This behaviour can be detected when scientists become unreasonably
over-confident that their theories are {\it true}, in the sense
that, in their opinion, Nature does follow them. Besides, those
who are prone to this kind of behaviour frequently do not accept any
challenge to their way of thinking. That makes matters even worse,
since they may reinforce the conservatism present at some particular
moment by helping to turn a healthy skepticism towards new
observations that challenge established ideas into an out of hand
rejection~of~them.

When a situation like that takes place, it creates an environment
where, in the view of those described above, the theory considered
as the best ``realization'' of the researched object assumes the role
of the supreme and only truth, never to be questioned. In these days,
however, science has become very dynamic and the theory elected
to be {\it the} true representation of Nature in some particular
area, and at some particular time, can be quickly toppled from
its position due to the unexpected arrival and imposing character
of new data or new discoveries. Then, if the dogmatic attitude remains,
what happens next is the urgent search and eventual replacement
of the toppled theory by a new ``supreme theoretical
truth'' which then becomes the 
dogma of the day.

The conservatism in the process described above can throw
the scientific community into deep confusion because, if the
community is large enough, opinions may be different among the
different research groups in the search for the new dogma. Then
one may expect a disagreement about the choice of ``the best''
theory, with the different groups choosing different theories.
So, we end up having a conflict of dogma within the community rather
than a scientific debate, which in practice becomes marginalized
or may even cease to exist.

One may therefore define {\it scientific dogmatism} as being
the unreasonably and unjustified over-confidence in certain theory,
over-confidence which stems from the misleading, and often unconscious, 
identification of the researched object with its correspondent
theory. Such identification implicitly or explicitly assumes
that the role of a scientific theory is to reveal Nature itself to us.
Such dogmatism causes over-confident researchers
to deliberately refuse as scientifically valid any theoretical
pictures different from theirs since, in their view, those different
theoretical pictures would ``contradict Nature itself''. As we
shall see, Boltzmann's epistemological theses are particularly
useful in clarifying this question. While they preserve the
freedom of personal choice in the creative theoretical work, they
also deny the notion that we can ever achieve an ultimate knowledge
of any scientific question, that is, they deny that we can reach
Nature itself, since in Boltzmann's view {\it any theory is
nothing more than a representation (image) of Nature} \cite{bo2,bbo2}. 

\section{\protect\normalsize \bf SCIENTIFIC THEORIES AS
         REPRESENTATIONS OF NATURE}

As stated above, at the end of the last century Boltzmann was
engaged in a passionate defense of his viewpoints, where he
sought to show that all scientific theories are nothing more
than representations of the natural phenomena. By being a
representation, a scientific theory cannot aim to know
Nature itself, knowledge which would explain why the natural
phenomena show themselves to us the way we observe
them, since such ultimate knowledge is, and will ever be,
unknowable. As a consequence, a scientific theory will never be
complete or definitively true. This viewpoint actually redefines
the concept of scientific truth by advancing the notion that it
is impossible to identify theory with the researched objects
since scientific theories are nothing more than images of Nature
(see \S \ref{verdade} below). In other words, a scientific
theory can, one day, be replaced by another. It is the possibility
of the replacement of one theory by another that defines and
constitutes the scientific progress, and that is diametrically
opposed to dogmatism~\cite{B1,BB1}.

Boltzmann's ideas about scientific models as representations are
clearly stated in the passage below, quoted from the entry {\sc model}
of the 1902 edition of the {\it Encyclopedia Britannica}:
\begin{quotation}
 \small \sf
 \noindent Models in the mathematical, physical and mechanical
 sciences are of the greatest importance. Long ago philosophy
 perceived the essence of our process of thought to lie in the
 fact that we attach to the various real objects around us
 particular physical attributes -- our concepts -- and by means 
 of these try to represent the objects to our minds. Such views
 were formerly regarded by mathematicians and physicists as nothing
 more than unfertile speculations, but in more recent times they
 have been brought by J.\ C.\ Maxwell, H.\ v.\ Helmholtz, E.\
 Mach, H.\ Hertz and many others into intimate relation with the
 whole body of mathematical and physical theory. On this view our
 thoughts stand to things in the same relation as models to the
 objects they represent. The essence of the process is the
 attachment of one concept having a definite content to each thing,
 but without implying complete similarity between thing and thought;
 for naturally we can know but little of the resemblance of our
 thoughts to the things to which we attach them. What resemblance
 there is lies principally in the nature of the connexion, the
 correlation being analogous to that which obtains between thought
 and language, language and writing. (\ldots) Here, of course, the
 symbolization of the thing is the important point, though, where
 feasible, the utmost possible correspondence is sought between
 the two (\ldots) we are simply extending and continuing the
 principle by means of which we comprehend objects in thought and
 represent them in language or writing \cite[p.\ 213]{model}.
\end{quotation}

It should be noted that the idea that scientific theories are
representations is still being echoed today, and an example of a recent
discussion can be found in \cite{bill}.

\subsection{\protect\bfit Theoretical Pluralism}

The most important epistemological conclusion which was reached by
Boltzmann from his debate against Ostwald's energeticism,
and which constitutes the core of his philosophical thinking,
is usually called theoretical pluralism. This is a consequence
of the thesis that all scientific theories are representations of
Nature. By being a representation, a scientific theory is, therefore,
initially a free creation of the scientist who can formulate
it from a purely personal perspective, where metaphysical
presuppositions, theoretical options, preferences for a certain
type of mathematical language, and even the dismissal of some
observational data, can enter into its formulation. All that in
the period where the theory is formulated. However, in order to
make this theory eligible to become part of science, it is
necessary for it to be confronted by the experience
\cite[\S 16, p.\ 286]{bo3}, \cite[p.\ 107]{bbo3}. If it is not
approved in this crucial test, the theory must be reformulated,
or even put aside \cite[p.\ 286]{bo3}, \cite[p.\ 225-226]{bbo3}.
Boltzmann also emphasized that, since all scientific theories
are, to some extent, free creations of scientists, scientific
work is impossible without the use of theoretical concepts,
which originates from the fact that it is impossible the formulation
of any scientific theory simply from the mere observation of
natural phenomena.

The theoretical pluralism also states that the same natural
phenomenon can be explained through different theories. Still
according to Boltzmann, this possibility has its origins in the
fact that, as seen above, any theory is a representation, a
construction, an image of the natural world. And nothing more.
One cannot do science in any other way. Either it is a construction,
a representation, or the theory is not scientific \cite[p.\
173-176]{guto}, \cite[p.\ 216]{citebol1}, \cite[p.\ 91]{citebbol1}.
In Boltzmann's words,
\begin{quotation}
 \small \sf
 \noindent (\ldots) Hertz makes physicists properly aware of
 something philosophers had no doubt long since stated, namely
 that no theory can be objective, actually coinciding with
 nature, but rather that each theory is only a mental picture of
 phenomena, related to them as sign is to designatum.

 \noindent From this it follows that it cannot be our task to find an
 absolutely correct theory but rather a picture that is, as
 simple as possible and that represents phenomena as accurately
 as possible. One might even conceive of two quite different
 theories both equally simple and equally congruent with phenomena,
 which therefore in spite of their difference are equally correct.
 The assertion that a given theory is the only correct one can only
 express our subjective conviction that there could not be another
 equally simple and fitting image \cite[p.\ 90]{citebbol1}.
\end{quotation}

In summary, theoretical pluralism synthesizes the fact that,
since knowledge of Nature itself is impossible, a
theory can only be better than another, not truer in the
non-Boltzmannian sense (see \S \ref{verdade} below). It is the
necessary mechanism which prevents science from running the risk
of stagnation. Within this perspective, truth can only be
{\it provisional}, and is in fact an approximation achieved by
different means, that is, by different theoretical images
\cite[p.\ 273, \S 3]{bo3}, \cite[p.\ 115-116]{bbo3}.

When Boltzmann advanced theoretical pluralism, he also had
another goal: to establish a clear and unreachable limit
for dogmatism, that is, a limit which it could not surpass. 
Boltzmann believed that once theoretical pluralism were
accepted and embodied in research practice, it would not
allow that, once proposed, a theory could be excluded from the
scientific scenario. 

Boltzmann also pointed out that the thesis that a scientific
theory is a representation was not new. Kant, in the 18th century,
and Maxwell, one of the most important influences upon him in
the middle of the last century, had both defended similar
theses. Other contemporary physicists, like Hertz and Helmholtz,
shared similar views \cite[p.\ 206]{citebol1}, \cite[p.\
83]{citebbol1}. By remembering that others like Kant and Maxwell
had already expressed similar propositions, Boltzmann wished
to make sure that any theory or model would be continuously
perfected, without being excluded by any other ``tribunal''
than the experience \cite[chap.\ 4]{guto}. But, before we go
into the relationship between Boltzmann's theoretical pluralism
and modern cosmology, we need to discuss in more detail
Boltzmann's notion of scientific truth.

\subsection{\protect\bfit Boltzmann's Concept of
	    Truth}\label{verdade}

One of the main features of modern science is that since the
beginning of the modern scientific revolution with Galileo,
scientists began to define truth as the {\it correspondence}
between models and observations. Nevertheless, since Boltzmann's
theses state that all scientific theories are representations of
natural phenomena, that is, they are not capable of determining
what {\it really} constitutes Nature, the concept of truth in
modern science should no longer be one which seeks to determine
Nature itself. Therefore, within the context of Boltzmann's
epistemological thinking, this concept of correspondence as
scientific truth becomes outdated as Boltzmann's views are
based upon the principle of theoretical pluralism. As a consequence,
since more than one model, or theory, may well represent the same
group of natural phenomena and/or experimental data, how it is
possible that scientists can choose one model among the possible ones?

At this moment Boltzmann advances another definition of scientific
truth: the {\it adequacy}. According to him, theory A is more
adequate than theory B if the former is capable of explaining more
rationally, more intelligibly, a certain set of natural phenomena, than
the latter. In his own words,

\begin{quotation}
 \small \sf
 \noindent (\ldots) let me choose as goal of the present talk
 not just kinetic molecular theory but a largely specialized
 branch of it. Far from wishing to deny that this contains
 hypothetical elements, I must declare that branch to be a picture
 that boldly transcends pure facts of observation, and yet I regard
 it as not unworthy of discussion at this point; a measure of my
 confidence in the utility of the hypotheses as soon as they throw
 new light on certain peculiar features of the observed facts,
 representing their interrelation with a clarity unattainable by
 other means. Of course we shall always have to remember that we
 are dealing with hypotheses capable and needful of constant further
 development and to be abandoned only when all the relations they
 represent can be understood even more clearly in some other way 
 \cite[p.\ 163]{blog}.

 \noindent (\ldots) We must not aspire to derive nature from our concepts,
 but must adapt the latter to the former. We must not think that
 everything can be arranged according to our categories or that 
 there is such a thing as a most perfect arrangement: it will only
 ever be a variable one, merely adapted to current needs \cite[p.\
 166]{blog}.
\end{quotation}

He also noted that since theories are images of Nature, all
have some explanatory power, and that a good theory is achieved
by being carefully crafted by scientists, in a process similar
to Darwin's Natural Selection:

\begin{quotation}
 \small \sf
 \noindent Mach himself has ingeniously discussed the fact
 that no theory is absolutely true, and equally hardly any
 absolutely false either, but each must gradually be perfected,
 as organisms must according to Darwin's theory. By being
 strongly attacked, a theory can gradually shed inappropriate
 elements while the appropriate residue remains \cite[p.\
 153]{citebol2}.
\end{quotation}

Once more, one should note that these ideas are still being
echoed today (for instance, see \cite[p.\ 214]{bill}).

\subsection{\protect\bfit The Search for a Good Theory}

It is important to stress that although theories are representations,
and, as we saw above, personal theoretical options can enter in
their formulation, they are {\it not} entirely arbitrary. The basic aim of
any theory is to represent something that is going on in Nature,
and a successful theory does achieve this to a considerable extent.
Therefore, such a theory can use some symbols, or a specific
mathematical language, just as conventions. However, since Nature
itself must be represented in it, conventions will always be limited
to only those aspects of the model, of the representation, which are
not perceived, in that theory, as being directly dictated by Nature.
Thus, under Boltzmann's perspective, one cannot say that theories
are just conventions, because after being carefully crafted by the
scientists as representations of unique, non-arbitrary, natural
phenomena, they become attached to them, and end up
saying something unique about what is going on in Nature.

Finally, besides being a good representation, there is still
another criterion capable of conducting the preference of
scientists towards one particular model: its predictive ability.
This is important because once a certain theoretical prediction
is confirmed, the scientific knowledge about Nature increases
quantitatively. A correct prediction is also important because
it is formulated within the context of a specific theoretical
picture. So, by being capable of predicting unknown phenomena,
a model shows all its explanatory power since it is not only
capable of announcing the already known ``pieces'', but it is also
able to go even further, showing the existence of other still
missing pieces which are necessary for a deeper and more organized
understanding of Nature. One cannot forget that one of the most 
important aims of science is to increase and organize our knowledge
about Nature, and thus, a certain theory is richer than others if
it is able to better contribute to such an increase and organization.
Such a preference for the richer theories makes them more likely
to be used, and developed, than the poorer ones, and after a
while the distance between them may be so great that it may no
longer be worth for researchers to keep on working with the
poorer representations, which are then put aside and, usually,
forgotten.

\section{\protect\normalsize \bf COSMOLOGICAL MODELS AS
	     IMAGES OF THE UNIVERSE}

In order to relate the previous discussion to issues in cosmology
we need, first of all, to distinguish Nature from its
representations, that is, from our theoretical images. Therefore,
we shall adopt the following difference between the terms `Universe'
and `universe'. The first term, with capital `U', will refer to the
aspects of Nature from which the different theoretical models are
built, while the second one, with small `u', will refer to the
models themselves. By means of this distinction, and bearing in mind
Boltzmann's theses, we can state that the different theoretical models
of the Universe are then ``universe models'', that is, ``cosmological
models'', or simply ``cosmologies'' or ``universes''. Cosmology is
then the science that attempts to create working representations
of the Universe.\footnote{ \ The terms Universe, universe,
cosmologies, etc, are found in the literature as usually having
meanings close to the ones given in this article. Therefore, what
is being done here is only to state precisely their meanings
in this context.} As a consequence, theoretical pluralism tells
us that there may be many different cosmologies, where each one
adopts different images of the Universe, although its true nature
is, and will always be, unknowable. 

From the perspective described in the previous sections we can
see how damaging dogmatic thinking can be, since in its most
basic sense it denies the cosmological community the option of
thinking differently than the current accepted view. That
inhibits the appearance of different theoretical representations,
which, according to the thesis of theoretical pluralism, are
fundamental for the development of modern cosmology. In other
words, dogmatism goes against theoretical pluralism. Therefore,
bearing in mind what we discussed in \S 1, it is not only
desirable, but {\it essential} that different theoretical
pictures, i.e., theoretical representations other than the
FLRW model emerge, be considered and developed in cosmology
nowadays. The present day dominance of the standard model,
due to its impressive achievements, should not be used as argument
against the emergence of models which may challenge the standard
FLRW picture. And if in the end those different representations
are wrong or produce worse models than the standard, cosmology
will gain in the process, especially the FLRW cosmologies.

One must stress that the theoretical pluralism does not
necessarily imply competition among the different theories, but
often means complementarity. This is exactly what Matravers, Ellis
and Stoeger \cite{mes} seek when they argue in favour of a more
phenomenological approach to cosmology. Inasmuch as, according to
Boltzmann's theses, all theories have some explanatory power,
all cosmologies end up saying something about the physical
process that are going on in the Universe, because not all
cosmologies use the same set of ideas and phenomena which
they seek to explain. Therefore, the emergence of different
cosmologies, far from being a problem for our better understanding
of the Universe, is essential for it. And if those different
cosmologies have elements which contradict each other, observations
provide the first mechanism, but not the only one, which allows
us to discard the inappropriate elements of the emergent
cosmologies.

The different cosmologies should either be in competition among
themselves or complement each other, but as none can be confused
with the Universe, no cosmology produces ultimate knowledge of it.
One cosmological model can only be provisionally better than another.
It is our observational interaction with the Universe that produces
the {\it empirical} basis upon which cosmologies are created, and
inasmuch as this interaction is basically technological, this
empirical basis will be changed by technological and theoretical
progress which itself creates the conditions for the partial or
complete transformation of the cosmological models. By the same
token, the diversity of technological means produces different
interactions and, therefore, different empirical basis that lead
to possible different cosmologies.

\section{\protect\normalsize \bf CONCLUSION}

From what we have seen in the previous sections, we can conclude
that dogmatism works against scientific progress and, to avoid it,
a change of attitude should be adopted by cosmologists. This
change is realized by the adoption of the theoretical pluralism
as a better epistemological framework for research in
cosmology. By means of this thesis it is possible to avoid the
deep rooted and unjustified belief in ideas which are nothing
more than just personal beliefs. As seen above, those ideas are necessary,
and perhaps indispensable, for the formulation of the different
representations of Nature, as such formulation is also a result of
the free creation of the scientists. However, those personal beliefs
will always be restricted to the theoretical models and, at best,
they can only generate better, or worse, representations of Nature.
Moreover, they should never be confused with the ``true'' Universe
since its ultimate reality is unknowable.

Inasmuch as no cosmological model can ever be capable of holding 
ultimate knowledge about the Universe, then it follows that no
cosmology can hold eternal truths. Then, the larger the number of
cosmologies available to cosmologists, the higher the chances of
we obtaining better representations of the Universe.

Scientific knowledge is best characterized by the continuous
search for better, but never definitive, representations of natural
phenomena. The replacement of a theory by another, the main feature
of modern science, can only happen if it is assured that no scientific 
theory can reach the stage of definitive truth. In other words,
a scientific theory can only be better than another one, and
nothing more. Therefore, since any cosmological model can only
aim to be a temporary explanation of what one chooses or is able to
observe and experiment in the Universe, when it is formulated it is
already doomed to disappear. The irony always present is that
no one can tell with precision when that will happen, unless
one takes a dogmatic posture.

In conclusion, we believe that it would be essential that
cosmologists in general recognize that their theories and models
about the Universe are nothing more than representations. This
explicit distinction is very important as it would create
the proper environment where different theories and models can
live together without the danger of dogmatism.

\newpage
\begin{flushleft}
{\normalsize \bf ACKNOWLEDGMENTS}
\end{flushleft}

We would like to express our thanks to M.\ A.\ H.\ MacCallum,
D.\ R.\ Matravers, S.\ F.\ Rutz, W.\ R.\ Stoeger and A.\ L.\ L.\
Videira for reading the preliminary manuscript and for useful
discussions, suggestions and remarks. The financial support
from CNPq is also acknowledged.



\begin{thebibliography}{99}
    \bibitem{d} Dingle, H., 1953. {\it M.\ N.\ R.\ astr.\ Soc.},
	  {\bf 113}, 393; quotation at page 394.
    \bibitem{mes} Matravers, D.R., Ellis, G.F.R. \& Stoeger, W.R.,
	  1995. {\it Q.\ J.\ R.\ astr.\ Soc.}, {\bf 36}, 29.
    \bibitem{tolman} Tolman, R.C., 1934. {\it Proc.\ Nat.\ Acad.\ Sci.\
          (Wash.)}, {\bf 20}, 169; see page 176.
    \bibitem{mm1} MacCallum, M.A.H., 1987. In {\it Theory and
	  Observational Limits in Cosmology}, Proc.\ Vatican Obs.\
	  Conf., p.\ 121, ed Stoeger, W.R., Specola Vaticana,
	  Vatican; see pages 121-127.
    \bibitem{wesson} Wesson, P.S., 1987. In {\it Theory and Observational
	  Limits in Cosmology}, Proc.\ Vatican Obs.\ Conf., p.\ 559,
	  ed Stoeger, W.R., Specola Vaticana, Vatican; see page 561.
    \bibitem{re} Rothman, T. \& Ellis, G.F.R., 1987. {\it Astronomy},
	  {\bf 15}, No.\ 2, 6; see page 22.
    \bibitem{k} Krasi\'nski, A., 1997. {\it Inhomogeneous Cosmological
          Models}, Cambridge University Press; see part IX, Afterthoughts. 
    \bibitem{vau} de Vaucouleurs, G., 1970. {\it Science}, {\bf 167},
	  1203; see pages 1203-1205.
    \bibitem{B} Boltzmann, L., 1905. {\it Popul\"{a}re Schriften},
	  J.A. Barth, Leipzig; English translations in \protect\cite{BB}.
    \bibitem{BB} Boltzmann, L., 1974. {\it Theoretical Physics and
	  Philosophical Problems: Selected Writings}, ed McGuinness, B.,
	  Reidel, Dordretcht; it contains the English translation of many
	  texts of \protect\cite{B}.
    \bibitem{B1} Boltzmann, L., 1897. {\it {\"{U}}ber die
	  Unentbehrlichkeit der Atomistik in der Naturwissenschaft},
          in \protect\cite{B}, pp.\ 141-157.
    \bibitem{BB1} Boltzmann, L., 1897. {\it On the Indispensability of 
	  Atomism in Natural Science}, \protect\cite{BB}, pp.\ 41-53.
    \bibitem{o} Ostwald, W., 1904. {\it Die Ueberwindung des
          wissenschaftlichen Materialismus}, in {\it In Abhandlungen
	  und Vortraege, Allgemeines Inhaltes (1887-1903)}, pp.\ 220-240,
	  Veit. and Co., Leipzig.
    \bibitem{guto} Videira, A.A.P., 1992. {\it Atomisme
	  Epist\'{e}mologique et Pluralisme Th\'{e}orique dans la
	  Pens\'{e}e de Boltzmann}, PhD thesis, Equipe Rehseis (CNRS),
	  University of Paris VII.
    \bibitem{debate} Videira, A.A.P., 1995. {\it Atomism and Energetics at
          the End of the 19th Century: the Luebeck Meeting of 1895},
	  preprint CBPF-CS-003/95.
    \bibitem{bo2} Boltzmann, L., 1890. {\it {\"{U}}ber die Bedeutung von 
          Theorien}, in \protect\cite{B}, pp.\ 76-80.
    \bibitem{bbo2} Boltzmann, L., 1890. {\it On the Significance of
          Theories}, in \protect\cite{BB}, pp.\ 33-36.
    \bibitem{model} Boltzmann, L., 1902. {\it Model}, in \protect\cite{BB},
          pp.\ 213-220.
    \bibitem{bill} Stoeger, W.R., 1993. {\it Contemporary Physics and
	  Ontological Status of the Laws of Nature}, in {\it Quantum
	  Cosmology and the Laws of Nature}, pp.\ 209-234, eds Russel,
	  R.J., Murphy, N., and Isham, C.J., Specola Vaticana.
    \bibitem{bo3} Boltzmann, L., 1899. {\it Die Grundprinzipien und
          Grundgleichungen der Mechanik, zweite Vorlesung}, in
	  \protect\cite{B}, pp.\ 270-289.
    \bibitem{bbo3} Boltzmann, L., 1899. {\it On the Fundamental Principles
	  and Equations of Mechanics}, in \protect\cite{BB}, pp.\ 101-128.
    \bibitem{citebol1} Boltzmann, L., 1899. {\it Ueber die Entwicklung
          der Methoden der theoretischen Physik in  neuerer Zeit}, in
          \protect\cite{B}, pp.\ 198-227.
    \bibitem{citebbol1} Boltzmann, L., 1899. {\it On the Development of
	  the Methods of Theoretical Physics in Recent Times}, in
	  \protect\cite{BB}, pp.\ 77-100.
    \bibitem{blog} Boltzmann, L., 1904. {\it On Statistical Mechanics},
	  in \protect\cite{BB}, pp.\ 159-172.
    \bibitem{citebol2} Boltzmann, L., 1903. {\it An Inaugural Lecture
	  on Natural Philosophy}, in \protect\cite{BB}, pp.\ 153-158.
\end{thebibliography}
\end{document}